\documentclass[pre,aps,twocolumn,showpacs,superscriptaddress,floatfix]{revtex4}
\usepackage{bm}
\usepackage{epsfig}
\usepackage{amsmath}
\usepackage{amsfonts}

\newcommand{\tw}{t_\mathrm{w}}
\newcommand{\ti}{t_\mathrm{i}} 
\newcommand{\tf}{t_\mathrm{f}} 
\newcommand{\Eb}{\varepsilon_\mathrm{b}}
\newcommand{\C}{\mathcal{C}}
\newcommand{\Cf}{\mathcal{C}_\mathrm{f}}
\newcommand{\Ci}{\mathcal{C}_\mathrm{i}}
\newcommand{\Cbar}{\overline{\mathcal{C}}}

\newcommand{\Cbari}{\overline{\mathcal{C}}_\mathrm{i}}

\begin{document}

\title{Fluctuation-dissipation ratios in the dynamics of
self-assembly}
\author{Robert L. Jack}
\affiliation{Department of Chemistry, University of California at Berkeley, Berkeley, CA 94709}
\author{Michael F. Hagan}
\affiliation{Department of Chemistry, University of California at Berkeley, Berkeley, CA 94709}
\affiliation{Department of Physics, Brandeis University, Waltham
 MA 02454}
\author{David Chandler}
\affiliation{Department of Chemistry, University of California at Berkeley, Berkeley, CA 94709}
\begin{abstract}
We consider two seemingly
very different self-assembly processes: formation 
of viral capsids, and crystallization of sticky discs.  At low
temperatures, assembly is ineffective,
since there are many metastable disordered states, which
are a source of kinetic frustration.
We use fluctuation-dissipation ratios to extract information about
the degree of this frustration.  We show that
our analysis is a useful indicator of the long term fate
of the system, based on the early stages of assembly.
\end{abstract}
\maketitle

\section{Introduction}

Self-assembly processes can be loosely defined as those
in which simple building blocks assemble spontaneously into
highly ordered structures.  
Assembly is of vital importance in biology, where cells use dynamically
assembled protein structures to control the shapes of lipid
membranes~\cite{selfAss_actin,
selfAss_tubulin,selfAss_budding}, and the life cycle of viruses involves
spontaneous assembly of their protein coats~\cite{selfAss_virus_expt,
selfAss_virus_theory}.  Recently, self-assembly has also been used 
to develop nanostructured materials~\cite{selfAss_persp,selfAss_general}, 
which often draw inspiration from biological systems.
The self-assembly of viral capsids~\cite{capsid}
has been the subject of elegant experimental~\cite{selfAss_virus_expt}
and theoretical studies~\cite{selfAss_virus_theory,HC,Louis06}.
In this article, we study a model~\cite{HC} designed to mimic
this assembly process.
At low temperatures, assembly is frustrated by the 
presence of long-lived disordered states.  
The avoidance of this frustration is crucial for successful assembly.
This effect is rather general, as we illustrate by also considering
the formation of ordered structures in 
a two-dimensional system of sticky discs.  
We analyze the crossover between frustrated and unfrustrated regimes,
and show that fluctuation-dissipation ratios (FDRs)~\cite{FDRtheory,
Jack-pq-fdr,fdr_obs,neg_fdr} associated with the early
stages of assembly are correlated
with the long-time yield of these processes.
This represents a new application of FDRs,
which have been discussed previously in the context of
glassy dynamics.  We discuss how this analysis might be
helpful in the design of self-assembling systems.

\begin{figure}[!ht]
\epsfig{file=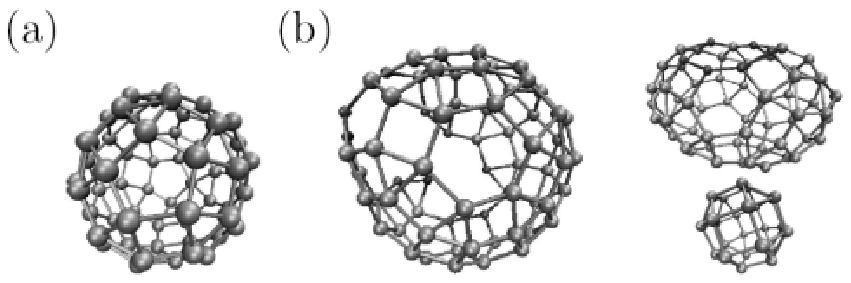,width=8.5cm}\par
\epsfig{file=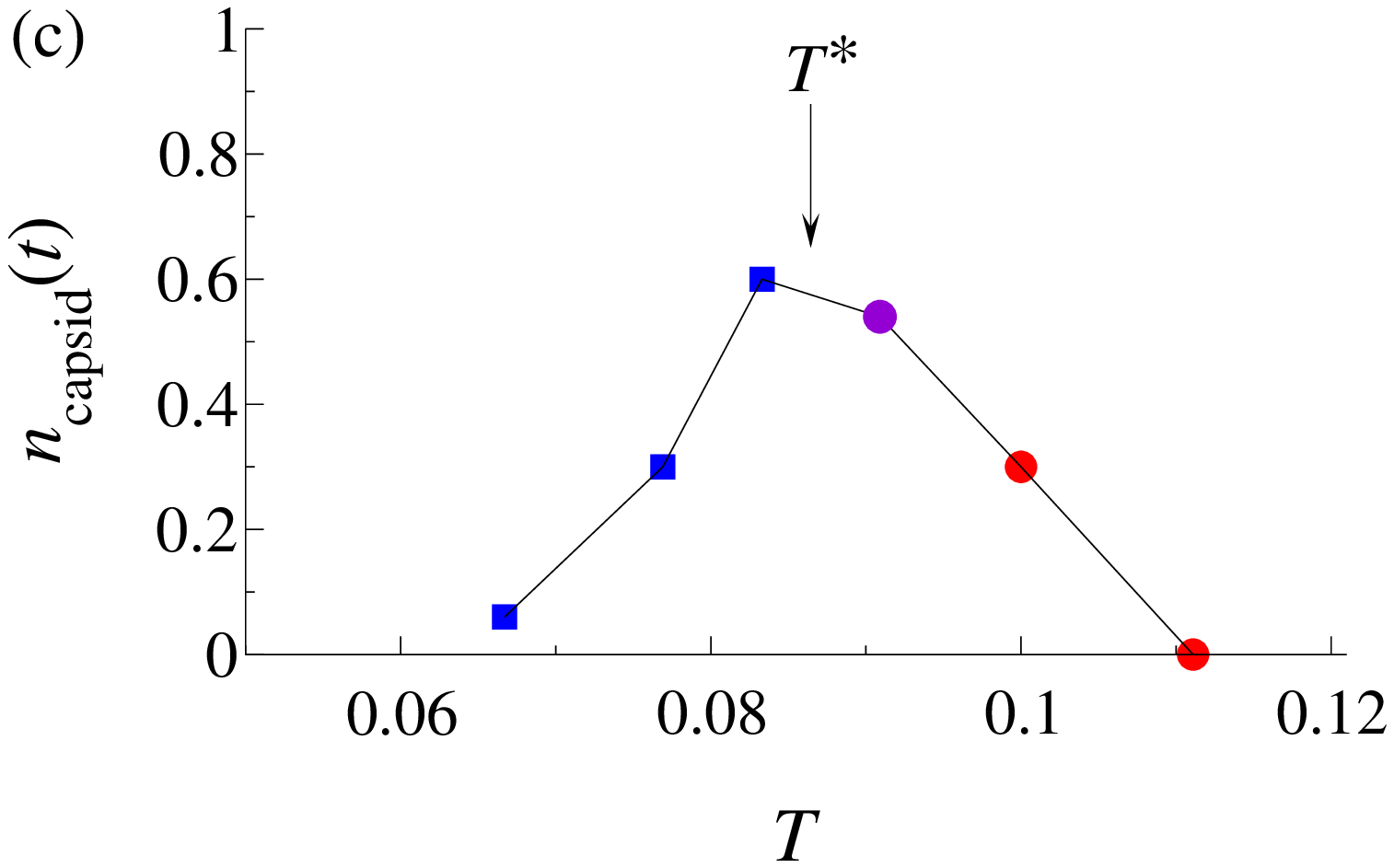,width=6.0cm}
\caption{(Color online)
Assembly of model capsids in the $\mathrm{B}_4$ model
of Ref. \cite{HC}.
(a)~A well-formed model capsid, with icosahedral symmetry.
(b)~Representative selection
of metastable states formed at reduced 
temperature $T=0.067$ and reduced time $t=3\times10^5$
(see the text for definitions).
(c)~Plot of the
the capsid yield at $t=3\times10^5$, which is non-monotonic
in the reduced temperature.
The yield is the fraction of particles in complete capsids,
identified as in \cite{HC}. Here and throughout,
red and blue symbols identify
high and low temperatures respectively. We also indicate
the approximate location of
the kinetic crossover, at reduced temperature $T^*$.
}
\label{fig:yield}
\end{figure}

In general, successful self-assembly requires both that the equilibrium
state of the system is an ordered structure, and that
the system reaches this ordered state in the time available for
the biological or experimental application.  
The first condition is thermodynamic: for the systems studied
here, the low energy ordered states are known, and this crossover
can be estimated by free energy arguments, as in \cite{HC}.
(Note however, that if `liquid-like' states are relevant
near the thermodynamic crossover, then this 
will lead to more complicated behavior, as in \cite{Louis06}.)
The second condition for successful assembly is kinetic in origin: 
it is illustrated
for a model system of viral capsid assembly in Fig.~\ref{fig:yield}. 
The degree of assembly shows a maximum at
a finite temperature $T^*$.  As the temperature is lowered through 
$T^*$, the ordered state becomes more probable at equilibrium, 
but the self-assembly process becomes less and less effective:
we refer to this change as a ``kinetic crossover''. 

\begin{figure*}
\epsfig{file=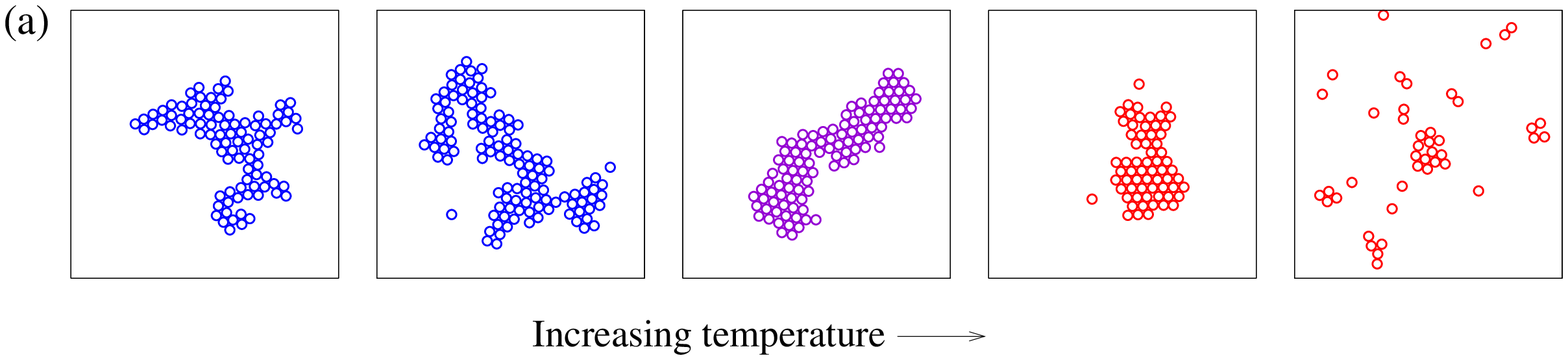,width=12cm}
\epsfig{file=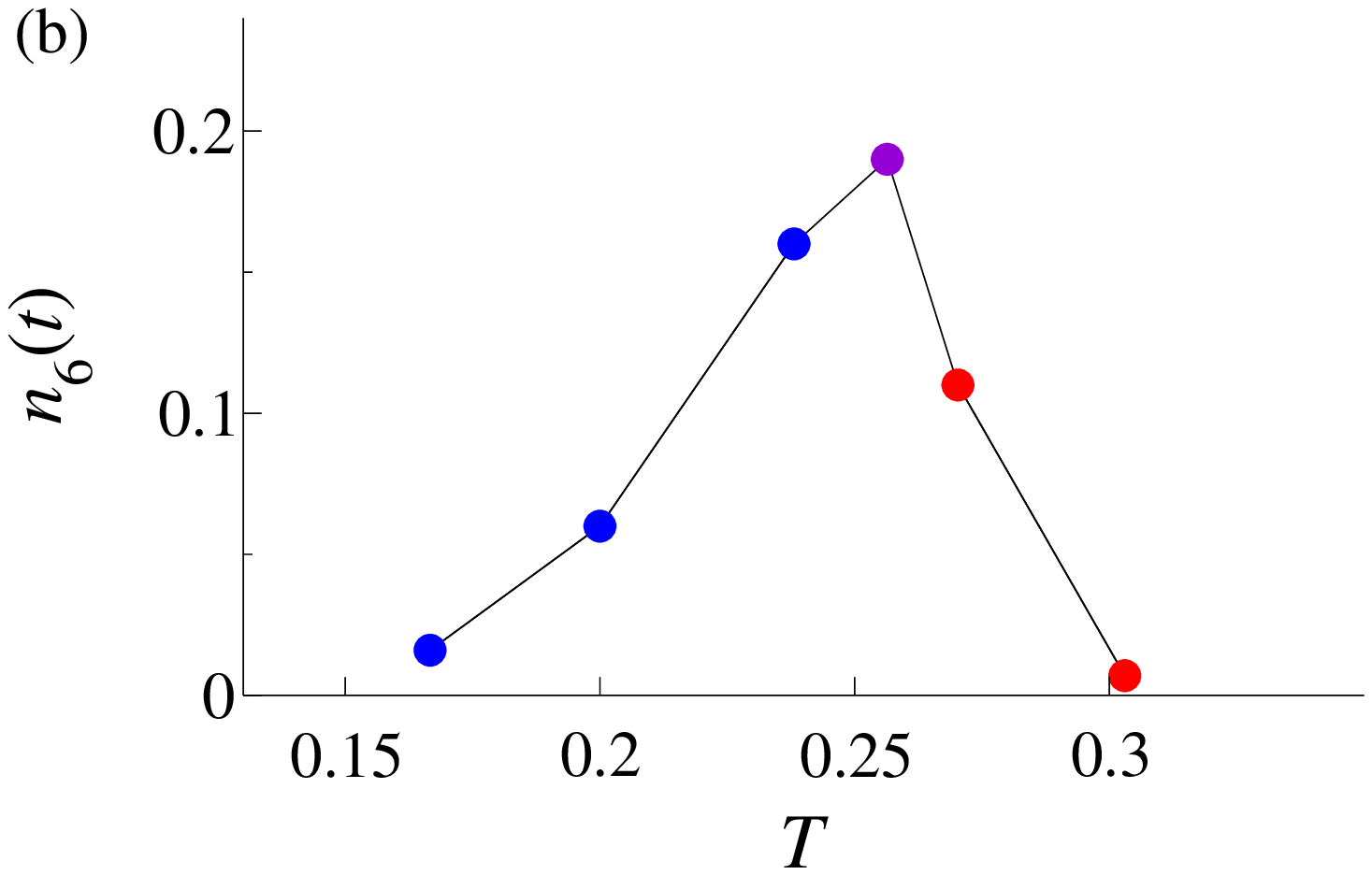,width=5.6cm}
\caption{(Color online)
Assembly of sticky discs.
(a)~Typical part-assembled structures 
at reduced time $t=5\times10^6$, and reduced temperatures 
$T=0.17,0.2,0.26,0.27,0.33$, from left to right.  Illustrated
regions are of size $25a_0\times25a_0$.  The crystallinity
is poor at low temperatures, due to
the metastability of the disordered states. 
(b)~Plot of the fraction of particles with 6 bonds, which is a
measure of the yield of the assembly process.  Compare
Fig.~\ref{fig:yield}c.
}
\label{fig:disc_yield}
\end{figure*}

The purpose of this article is to use dynamical observables
to study the behavior near $T^*$.  Since this is the regime
of most efficient assembly, it is relevant both
biologically and for applications of self-assembly in
nanoscience.  While the kinetic crossover can always be identified
by measuring the degree of assembly, as in Fig.~\ref{fig:yield},
achieving this in a computer simulation
requires access to long timescales, which restricts the range of 
systems that can be studied.  In this article, we show 
how FDRs can be used to locate the
kinetic crossover using  
simulations on relatively short time scales. (It is 
necessary to average over many such short simulations, but such 
averaging is trivially parallelizable.)  
We also discuss how these response functions might be measured
experimentally in ordering processes that occur on
complex energy landscapes. 

\section{Models}

\subsection{Model capsids}

The first model that we discuss describes the
assembly of viral capsids.  Full details are given in \cite{HC}.  
The model consists of rigid subunits, the ``capsomers'', which
interact by isotropic repulsive forces,
and directional attractions.  The low energy states in 
the model contain ``capsids'', each of which consists of 60 subunits in a 
cage structure, with icosahedral symmetry.
We use the $\mathrm{B}_4$ variant of this model, which means that the
attractive potential favors the capsid structure shown in
Fig.~\ref{fig:yield}a.
The subunit diameter is $\sigma$, and the 
density of subunits is $\rho$.
The parameters of the model are the reduced capsomer density 
$\rho\sigma^3$ and the reduced temperature 
$T$ (measured in units of $\Eb/k_\mathrm{B}$, where
$\Eb$ is the energy associated with the attractive potential
and $k_\mathrm{B}$ is Boltzmann's constant). 
In addition, the specificity
of the directional attractions is controlled by the angular parameters
$\theta_\mathrm{m}$ and $\phi_\mathrm{m}$.
The data of this article are obtained under the representative conditions
$\rho\sigma^3=0.11$, $\theta_\mathrm{m}=1.5$ and $\phi_\mathrm{m}=3.14$. 
We simulate a system
of 1000 capsomers in a cubic box with periodic boundaries.  The
capsomers evolve according to overdamped
Brownian dynamics, and the unit
of time is $(\sigma^2/48D)$, where $D$ is the
capsomer diffusion constant.  The rotational diffusion
constant of each capsomer is $D_\mathrm{r}=2.5(D/\sigma^2)$, 
as in \cite{HC}.

\subsection{Sticky discs}

We also consider a second model 
whose subunits are sticky discs which interact
by an attractive square-well potential of depth $J$ and range
$a_0$, and a repulsive hard core of range $0.9 a_0$.  We quench the system
into the solid-vapor phase coexistence regime, so that the equilibrium
state has most of the discs in a single close-packed crystallite.  However,
we use Monte Carlo dynamics that
are chosen to accentuate the effects of kinetic frustration.
We move bonded clusters as rigid bodies, allowing 
translation and rotation, but no internal rearrangements.  
To reflect the slow motion of large clusters, we use
an average translational step size of $0.1(a_0/M)$ and 
a rotational step of $\pi/(10I)$ radians, 
where $M$ is the number of particles in the cluster and
$Ia_0^2$ its moment of inertia (in units of the disc mass).
The reduced time $t$ is measured in Monte Carlo sweeps, and the
reduced temperature $T$ is measured in units of $J/k_\mathrm{B}$.
Clusters can rearrange only by bond breaking.
These events are sampled by the
`cleaving algorithm' of~\cite{WhitelamG05}, with equal
fictitious and physical temperatures.  It is an off-lattice
generalization of the Swendsen-Wang algorithm~\cite{SwendsenW}.
At low temperatures, the dynamics mean that bonds are broken 
very rarely, and aggregation of the discs is diffusion limited.
At $T=0$, the system reduces to diffusion-limited cluster
aggregation (DLCA)~\cite{dlca}.

The crossover from effective to ineffective assembly
in the capsid system was shown in Fig.~\ref{fig:yield}.
We show similar results for the disc system in 
Fig.~\ref{fig:disc_yield}.  The system contains
$400$ discs in a periodic square box of side $100a_0$.
The system does not reach full phase separation into 
close-packed structures on the time scales accessible to our simulation, 
so all of our data is in the out-of-equilibrium regime. 

\section{Fluctuation-dissipation ratios} 
\label{sec:fdr}

The non-monotonic yields shown in Figs.~\ref{fig:yield}c
and~\ref{fig:disc_yield}b mean that for the observation
times considered, and when
the temperature is small, reducing the temperature does
not result in a decrease in of the total energy.
This kind of `negative response' to temperature
perturbations is familiar in systems
with activated dynamics~\cite{neg_fdr}.  
In the self-assembling systems considered here, the kinetic
crossover at $T^*$ is associated with a change
from positive to negative response on the long time scales
considered in Figs.~\ref{fig:yield}c and~\ref{fig:disc_yield}b.  
In this section, we use fluctuation-dissipation
ratios (FDRs) to show that 
the crossover between positive and negative response
has signatures that can be detected on much shorter time
scales.

\subsection{Basic idea}

Fluctuation-dissipation ratios (sometimes also called
correlation-response ratios) have been widely studied in the context
of aging of glassy systems \cite{FDRtheory}.  Imagine applying an
instantaneous perturbation to a single subunit (disc
or capsomer) at a 
time $\tw$, and measuring the effect of this perturbation
at some later time $t$.  For a system at
equilibrium, the fluctuation-dissipation theorem (FDT) 
relates the response to small perturbations
to the relaxation of spontaneous fluctuations \cite{IMSM}.

In general, we can measure the 
fluctuations and responses of any observable.
Here, we focus on the the energy of a given subunit.
In both of our model systems, the total energy comes 
from interactions between pairs of subunits, 
$
  E_0=(1/2)\left.\sum'_{ij}\right. E_{ij},
$ 
where the primed sum excludes terms with $i=j$.  We denote the energy of
the $i$th monomer by 
\begin{equation}
  E_i\equiv(1/2)\sum_{j(\neq i)} E_{ij}.
\end{equation}

We measure the responses in the system as follows.
Starting from a given initial state, the system 
assembles for a `waiting time' $\tw$.  We then turn on a
perturbation to the energy:
$\delta E(t) = \sum_i h_i E_i \Theta(t-\tw)$, where $h_i$
is the (small) field applied to the $i$th subunit, and
$\Theta(x)$ is the unit-step function.  
We measure the integrated response to this field,
\begin{equation}
\chi(t,\tw) = \left(
              \frac{\partial \langle E_i(t) \rangle_{\tw}} 
                   {\partial (\beta h_i)} \right)_{\bm{h}=\bm{0}},
\label{equ:def_chi}
\end{equation}
where the notation $\bm{h}=(h_1,h_2,\dots)$, and
$\beta^{-1}$ is the temperature multiplied by 
Boltzmann's constant. 
The average is over trajectories of the system
in the presence of the perturbation.
In practice, we evaluate the partial derivative by assigning
$h_i=\delta h$ to half of the subunits (selected at random),
and $h_i=-\delta h$ to the other half.  In the
linear response regime (small $\delta h$), the mean energy
at $\bm{h}=\bm{0}$ can then
be estimated by $\overline{E}(t)=N^{-1} \sum_i E_i(t)$, and the response
by $\sum_{i} [E_i(t) - \overline{E}(t)]/h_i$.  These
quantities are then averaged over many independent runs
of the dynamics.  Our results for the capsid system were obtained
at $\delta h=0.05$ and those for the disc system were obtained
at $(\delta h/T)=0.3$.
These values are small enough that our estimates of $\chi(t,\tw)$
change very little if $\delta h$ is reduced, which
indicates that we are in the linear response regime.  For
systems with Monte Carlo dynamics, such as the disc
system, the derivative in
Eq.~(\ref{equ:def_chi})  
can also be evaluated as a correlation function for
the unperturbed dynamics, in which case it is no longer necessary
to apply the field $h_i$ directly~\cite{Berthier-nofield,Chatelain}.

For a system at equilibrium, the fluctuation-dissipation
theorem \cite{IMSM} states that
\begin{equation}
\chi(t,\tw) = C(t,t) - C(t,\tw)
\end{equation}
for all $t$ and $\tw$,
where
\begin{equation}
  C(t,\tw) \equiv \langle E_i(t) E_i(\tw) \rangle - \langle E_i(t) \rangle
\langle E_i(\tw) \rangle, 
\end{equation} 
Alternatively, we can define the response to an instantaneous
perturbation (impulse response), as a derivative of
the integrated response:
$R(t,\tw) = -\partial \chi(t,\tw) /\partial \tw$.  In that case,
the FDT states that
\begin{equation}
R(t,\tw) = \frac{\partial C(t,\tw)}{\partial \tw}.
\end{equation}

Away from equilibrium, we define 
the correlation-response ratio $X(t,\tw)$ by
\begin{equation}
R(t,\tw) =  X(t,\tw) 
\frac{\partial C(t,\tw)}{\partial \tw}.
\end{equation}
Thus, $X(t,\tw)$ is the response of the system to an instantaneous
perturbation, normalized by the response of an equilibrium system
with the same fluctuations.  

\begin{figure}
\epsfig{file=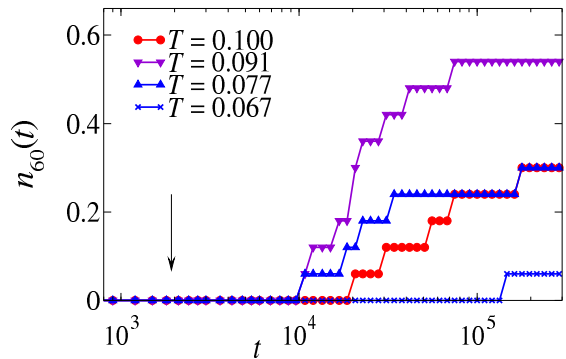,width=8.0cm}
\caption{(Color online)
 Sample trajectories in the capsid system, showing $n_{60}(t)$,
 defined as the fraction of particles in bonded clusters of size
 60.  We use a logarithmic scale for the reduced time $t$.
 The fraction $n_{60}(t)$ reflects the number of capsids in the
 system, since disordered clusters containing exactly 60 subunits
 are rare.  The first capsids appear at times around $10^4$.
 The system is away from global equilibrium until reduced
 times are at least
 of the order of $10^5$.  The arrow indicates the maximal time
 associated with our measurements of correlation and response
 functions (Figs.~\ref{fig:resp}-\ref{fig:fdr}).
}
\label{fig:times}
\end{figure}

In equilibrium, the fluctuation-dissipation theorem implies
that $X(t,\tw)=1$ for all $t$ and $\tw$.  Away from equilibrium,
$X(t,\tw)$ may take any value.  It is most conveniently obtained
from the gradient of a parametric plot of $\chi(t,\tw)$ against
$C(t,\tw)$, where the parametric variable is the waiting time 
$\tw$~\cite{FDfoot}.
We will see that parametric plots distinguish between systems
above the kinetic crossover region, and those below it.  This application
of the FDR is the main result of this article.

\subsection{Results}

\begin{figure}
\epsfig{file=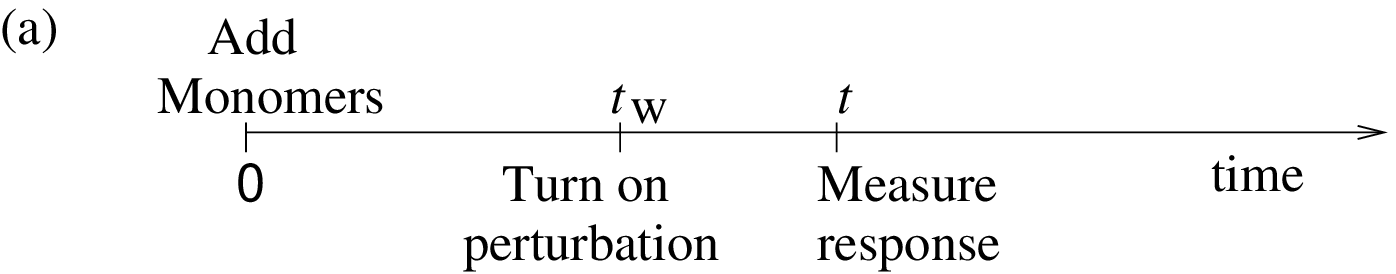,width=7.3cm}\par
\epsfig{file=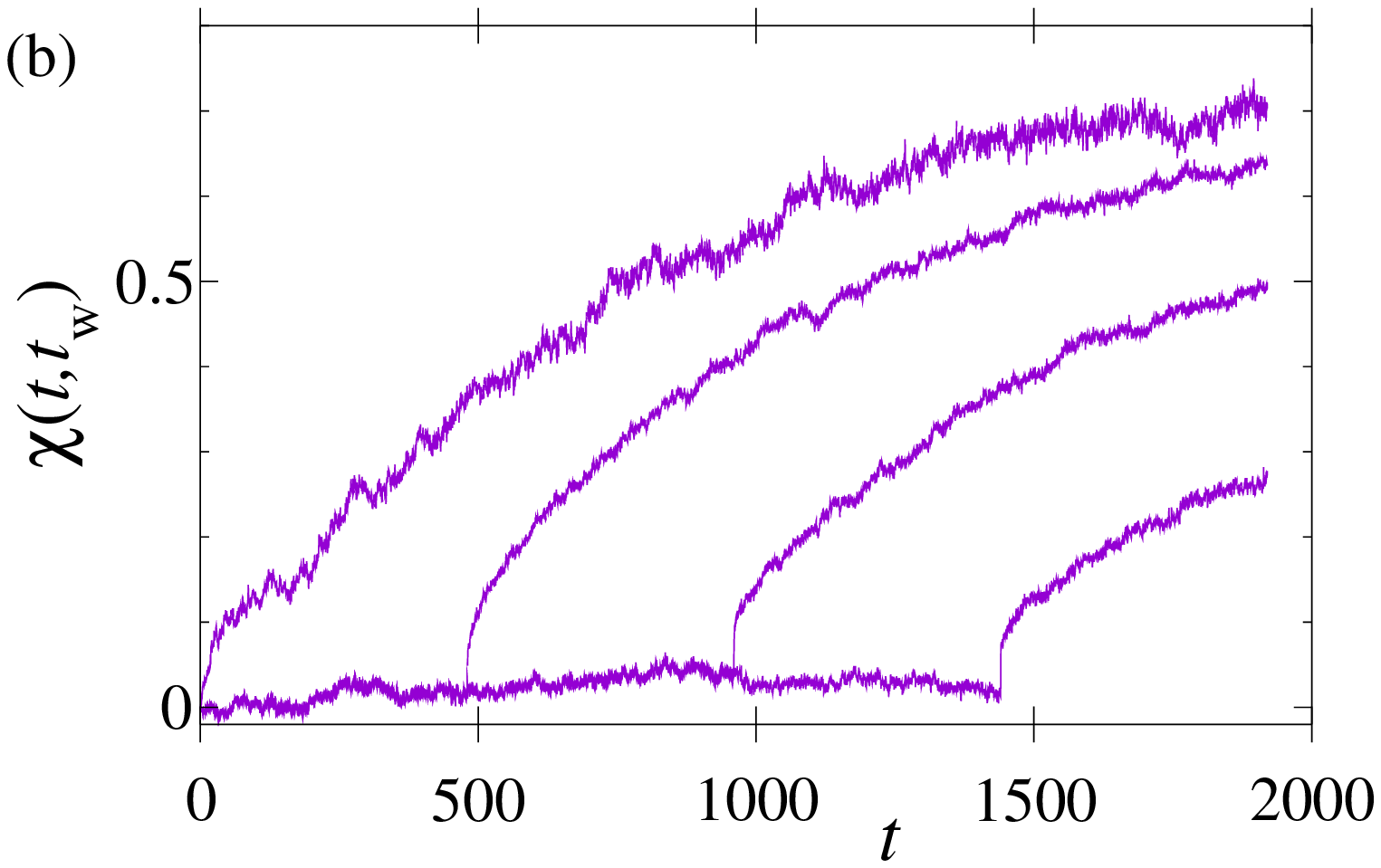,width=7.3cm}\par
\epsfig{file=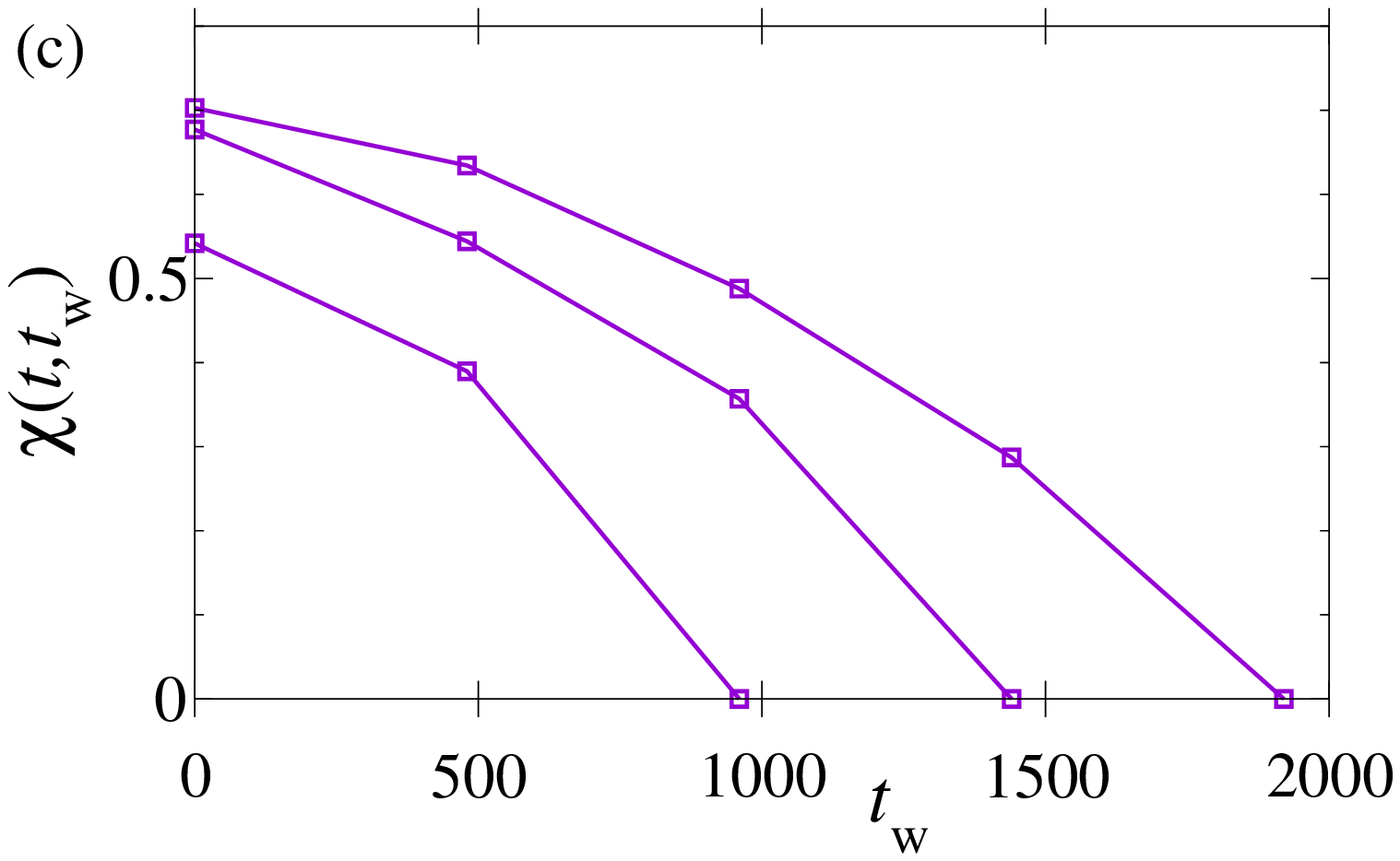,width=7.3cm}
\caption{(Color online)
(a) A time-line illustrating the simulation protocol used to 
measure the response.
(b) Response in the 
capsid system (in units of $\Eb^2$) 
at reduced temperature $T=0.091$,
as a function of time $t$, for $\tw=0,480,960,1440$.  The data
are plotted with lines, since each simulation yields data points
for all $t$.
(c)  Plot of the response as a function of waiting time $\tw$,
for $t=960,1440,1920$.  This is a replot of
some of the data of the middle panel, but it allows estimation
of the impulse response $\partial\chi(t,\tw)/\partial \tw$.  \
In this case, the
data are shown as points (squares), and points with the same value of $t$ are
connected by lines.
}
\label{fig:resp}
\end{figure}

\begin{figure}
\epsfig{file=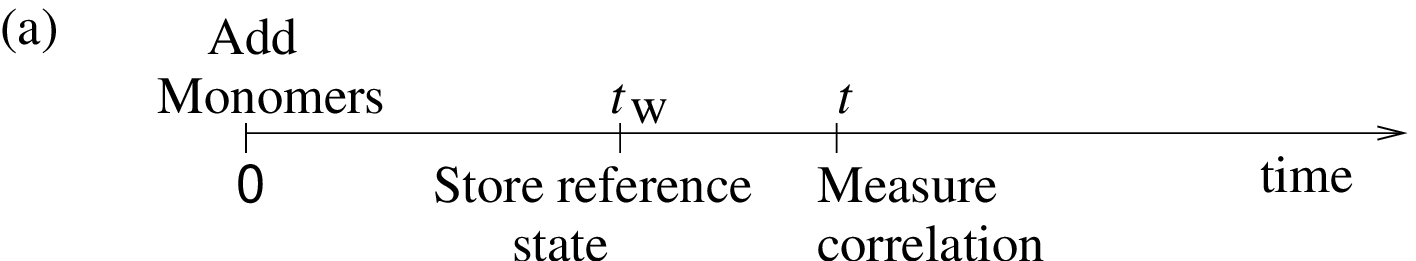,width=7.3cm}\par
\epsfig{file=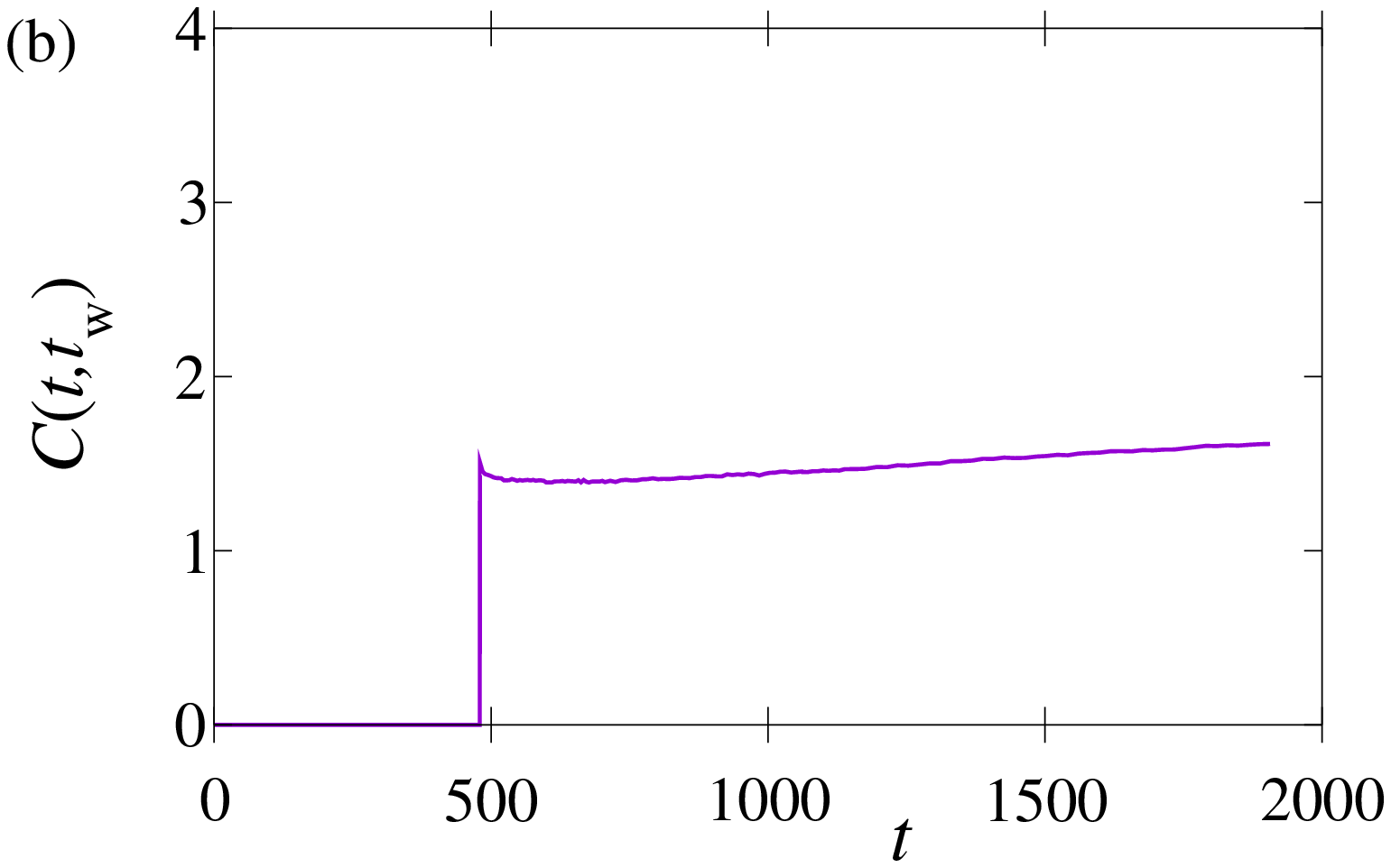,width=7.3cm}\par
\epsfig{file=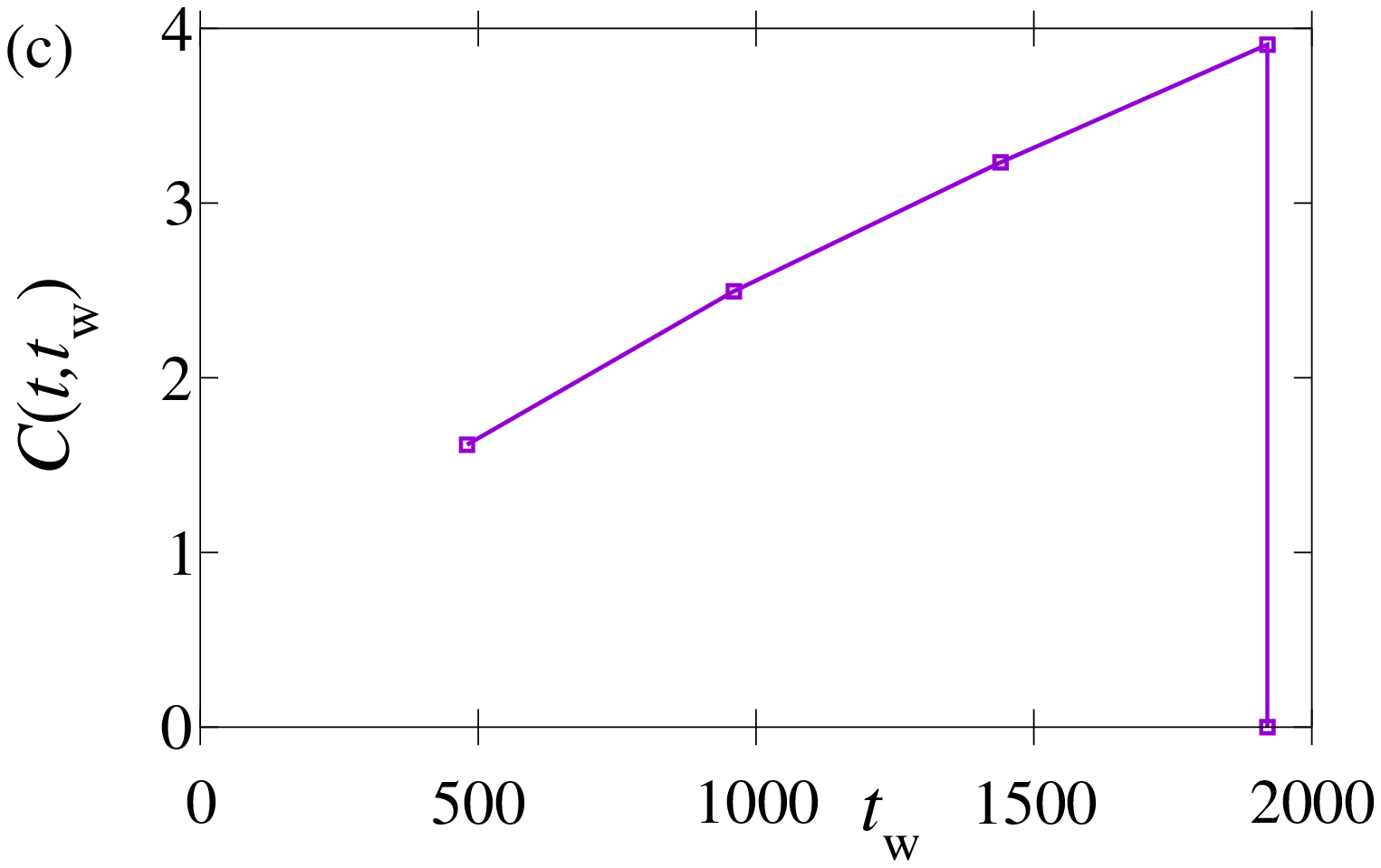,width=7.3cm}
\caption{(Color online)
(a) A time-line
indicating the simulation protocol used to measure the 
correlation.
(b) Correlation function in the 
capsid system (in units of $\Eb^2$) 
at $T=0.091$,
as a function of reduced time $t$, for $\tw=480$.
(c)  Correlation as a function of $\tw$, for $t=1920$.
The absence of time-reversal symmetry is clear.  
}
\label{fig:correl}
\end{figure}

\begin{figure}
\epsfig{file=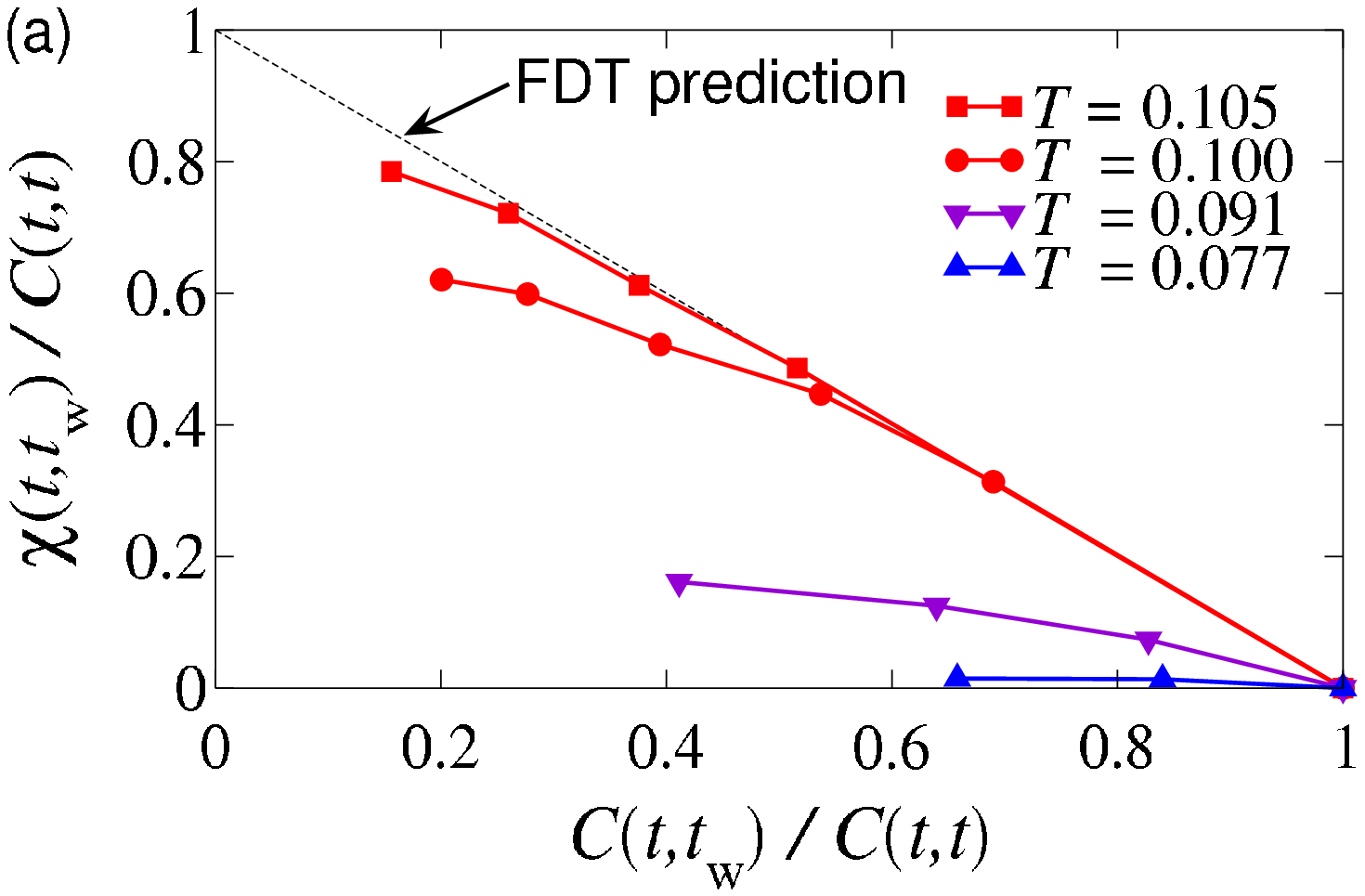,width=8cm}\par
\epsfig{file=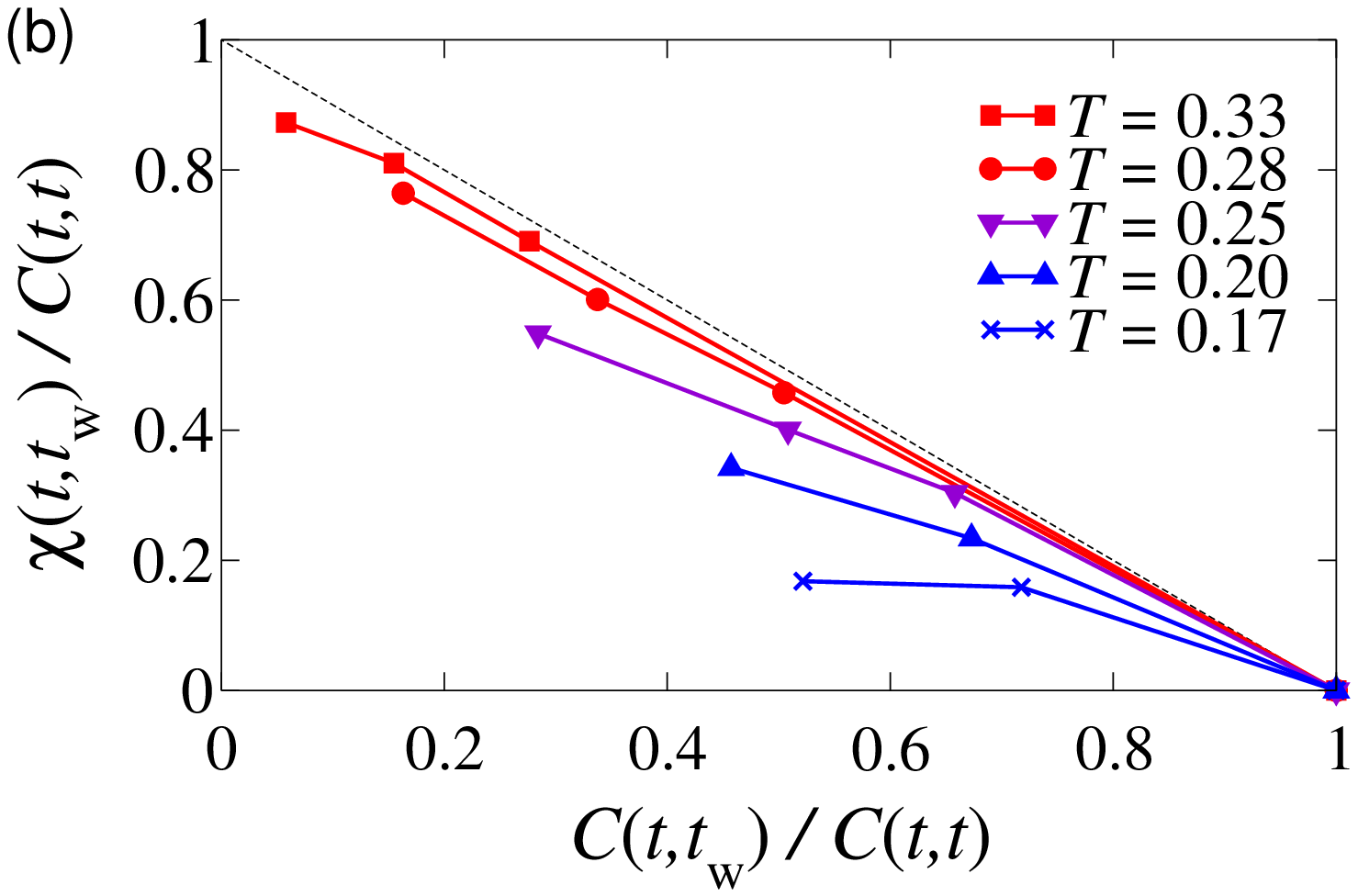,width=8cm}
\caption{(Color online)
Correlation response plots for (a)
the capsid system at $t=1920$, and $960<\tw<t$, and 
(b) the disc system at $t=8\times10^4$.
These systems are all well away from equilibrium,
but the response is in accordance with the prediction of FDT 
at the higher temperatures.  The response
decreases rapidly as the system passes through the kinetic
crossover and falls out of equilibrium.  
The red and blue coloring is consistent with that
of Figs.~\ref{fig:yield} and~\ref{fig:disc_yield}}
\label{fig:fdr}
\end{figure}

In Fig.~\ref{fig:times}, we illustrate the time scale
associated with capsid formation.  The first capsids form
in the system at times around $10^4$, and all systems shown are 
significantly out of equilibrium until reduced times 
at least as large as $10^5$.  
The yield measurements of Fig.~\ref{fig:yield}
were taken at $t=3\times10^5$.  
As time proceeds, the system evolves increasingly
slowly towards the equilibrium state.  We will show 
correlation and response data at times of order $10^3$,
so the system is still well away from global equilibrium in all cases.
However, we will find that systems at temperatures above
the kinetic crossover region have responses in accordance with FDT,
while those below it do not.  In the
disc system, the snapshots of Fig.~\ref{fig:disc_yield}  show
that the system is well away from equilibrium at times around $5\times10^6$.
For that system,
we will show correlation-response data at much earlier times, those less
than $10^5$.

Some results for the capsid response function are shown in 
Fig.~\ref{fig:resp}, where we show how multiple simulations 
are used to plot the response as a function of $\tw$ for fixed $t$, 
which is useful for estimating the impulse response.  
A typical correlation function is shown in Fig.~\ref{fig:correl}.

Results for the FDR in both capsid and disc systems
are shown in Fig.~\ref{fig:fdr}, where we have normalized
both correlation and response by the equal time fluctuation
$C(t,t)$.  [The function $C(t,t)$ is independent of $\tw$, so the 
gradient of the parametric plot is $-X(t,\tw)$, and is unaffected
by the normalization.]  
Above the kinetic crossover, assembly is taking place, but the
energy response is in accordance with FDT, so $X(t,\tw)\approx 1$,
at least for the times that we considered.
As the system passes through the kinetic crossover, the FDR
shrinks.  While it can be convenient to characterize
this crossover by the single temperature $T^*$,
it is more accurate to think of a temperature range over which the 
long-time behaviour of system changes smoothly from
effective to ineffective assembly.  This smooth
change is accompanied by a smooth change in the FDR.

We conclude that if a system is to be designed
so that it assembles effectively, the correlation-response
ratio can be used to obtain a general prediction for the 
regime of good assembly, before running the long
simulations required to test the yield directly.

Finally, note that we constructed Fig.~\ref{fig:fdr}
using data at constant $t$ and variable $\tw$, since the gradient
of this plot gives the FDR.  This procedure
requires a separate simulation for each data point.  However, if we only 
wish to know if the integrated response
is small or large compared to the FDT prediction, 
it is sufficient to use data at a single $\tw$: a simple comparison
of $C(t,t)-C(t,\tw)$ and $\chi(t,\tw)$ is already quite informative in
that case (note however \cite{FDfoot}).  

\section{Discussion}
\label{sec:discussion}

We now consider the kinetic and
thermodynamic crossovers in a little more detail.
We then discuss how the change in FDR at the kinetic
crossover arises, and the extent to which we
expect it to generalize to other self-assembling systems.

\subsection{Thermodynamic and kinetic crossovers}
\label{sec:xover}

We measure the yield of our assembly processes by running
long simulations of length $t_\mathrm{yield}$ (recall
Figs.~1 and 2).
These simulations have three types of final state.
At high temperatures,
no assembly takes place, and the system consists primarily of free
subunits.  At low temperatures, the system evolves into a state
that consists primarily of disordered metastable clusters.

We also find an intermediate
temperature regime, in which the final state has a substantial
quantity of assembled products.
This regime is delineated by two crossovers.
For an operational definition of the high temperature
crossover, we impose a threshold on the relative probabilities of
bonded and free subunits at time $t_\mathrm{yield}$.  While
this definition depends on $t_\mathrm{yield}$, the position
of the crossover has a well-defined limit as $t_\mathrm{yield}\to\infty$,
which can be evaluated from the contribution of free subunits
to the thermodynamic partition function of the system.  Thus
we refer to this crossover as ``thermodynamic''.  

To define the low temperature crossover, we consider the relative
probabilities of disordered clusters and correctly-assembled products
at $t_\mathrm{yield}$.  As the temperature is reduced,
the maximum of the yield occurs when the disordered clusters
are common enough to significantly impede assembly.
We therefore identify this maximum with the low temperature crossover.
If we calculate the yield in the equilibrium state, we expect it to depend
monotonically on the temperature, since the correctly-assembled states
minimise the total energy in both of our systems.  Thus, the presence
of the maximum in the yield is a kinetic effect, that arises from the slow
annealing of disordered clusters.  This motivates our use
of the term ``kinetic crossover''.   Clearly, the existence
of a regime of efficient assembly requires that the kinetic crossover
is not too close to the thermodynamic one.  If the system
crosses over smoothly from free subunits to disordered
clusters, then there is no temperature at which assembly
is efficient on the time scale $t_\mathrm{yield}$.

\subsection{Local equilibration}

We now return to the link between the kinetic crossover
and the FDR.  The general idea is that 
dynamics that is locally time-reversal symmetric allows
disordered states to anneal into ordered states.  This
idea is not new (for example, see Ref.~\cite{Whitesides02}, 
especially its Fig. 1).  However, the FDR provides a 
quantitative measure of this effect.

The crystalline state of the two-dimensional system of discs 
is close-packed, with each particle bonded to six neighbors.
During assembly, the fraction of such particles in a given cluster
provides a measure of its crystallinity.
As clusters form, there are many possible
states with low crystallinity, and the system tends to
visit these states quite frequently.  The effectiveness
of assembly depends on 
whether these states are able to anneal into 
crystalline clusters.  This annealing becomes more difficult 
as the disordered clusters grow.  For example, annealing the
disordered 
clusters of Fig.~\ref{fig:disc_yield} into crystallites
requires highly co-operative processes with large
activation energies, while annealing small disordered
clusters requires less co-operativity.  

Our results indicate that near optimal assembly, large
disordered
clusters are avoided because the system remains \emph{locally
equilibrated} at each stage of the assembly
process (although the system is globally out of equilibrium).
At any stage of assembly, there will be a set of likely states.  
The condition of local equilibration is that the relative
probabilities of these likely states reflect their
relative Boltzmann weights.  If this condition holds, the
system avoids the disordered states that are precursors
to the large disordered clusters of Fig.~\ref{fig:disc_yield}. 
For example, small disordered clusters have smaller Boltzmann 
weights than crystalline clusters of the same size, so local
equilibration suppresses the disordered states.  On the
other hand, if disordered
states are likely at any stage of assembly, this indicates
that they are not being annealed into crystallites, and
are likely to evolve into larger disordered clusters.

To link this argument with the FDR, we first
demonstrate a link between local equilibration
and an approximate time-reversal symmetry.  We consider
two states $\C$ and $\C'$ that are both likely at a given
stage of assembly.  The rate with which the system
makes transitions from $\C$ to $\C'$ is
\begin{equation}
\gamma(\C\to\C',t) = W(\C'|\C) p(\C,t) 
\end{equation}
where $p(\C,t)$ is the probability that the system is in state
$\C$ at time $t$, and $W(\C'|\C)$ is the rate for transitions 
to state $\C'$ given
that the system is initially in state $\C$.  [The rate
$W(\C|\C')$ depends only on the dynamical rules of the model,
while the rate $\gamma(\C\to\C',t)$ depends also on the state
of the system at time $t$].

For models that obey detailed balance, we have
\begin{equation}
W(\C'|\C) \exp(\beta E_{\C'})  = W(\C|\C') \exp(\beta E_{\C}),
\end{equation}
Further, if the system is locally equilibrated then
we have
\begin{equation}
p(\C,t) \exp(\beta E_{\C}) \approx p(\C',t) \exp(\beta E_{\C'}).
\end{equation}
where $\C$ and $\C'$ are likely states at this time.
Thus, the rates for forward and reverse transitions between
these states are equal:
\begin{equation}
\gamma(\C\to\C',t) \approx \gamma(\C'\to\C,t).
\end{equation}
This relation is an approximate 
time-reversal symmetry of the locally equilibrated state,
which holds on time scales for which the set of
likely states is not changing significantly. 

The extent to which this approximate time-reversal symmetry 
holds is correlated with the degree of local equilibration, and
hence with the extent to which the system is discriminating
between high-energy disordered states and low-energy ordered ones.
By avoiding the high-energy disordered states, the locally
equilibrated system tends to assemble effectively.

To link this local equilibration with the FDR,
we show in the appendix that, for systems obeying
detailed balance, deviations from FDT arise from
differences between the probabilities of trajectories and
their time-reversed counterparts, during the time between
perturbation and measurement.  The key result is (\ref{equ:fdr_rev}). 
We conclude that the FDR is a probe of the degree to
which the system obeys time-reversal symmetry between times
$\tw$ and $t$, and hence of the degree of local equilibration.

Thus, our results for both capsid and disc systems (Fig.~\ref{fig:fdr})
are consistent both with our hypothesis
that the system falls out of local equilibrium at the kinetic
crossover, and with our interpretation of the FDR as a measure
of local equilibration.  The parametric plots 
summarize the important features of the correlation and response functions,
in a single system-independent plot, in which time and energy scales are
rescaled away.  The qualitative similarities in the behavior of
the FDR are all the more remarkable given the different
dimensionalities of the two models that we consider, 
and the very different structures of their assembled states.

\subsection{Generic and non-generic features of the FDR}

While the behavior of both capsid and disc systems are
both consistent with our analysis above,
there are important differences between the
two panels of Fig.~\ref{fig:fdr}.  In particular,
at the peak of the assembly curve, the response in the disc
system is larger than the corresponding response in the capsid
system.  

The reason for this difference can be inferred from the states shown
in Fig.~\ref{fig:yield} and~\ref{fig:disc_yield}.  In the disc
system, the crystallinity of the product is rather low at all temperatures.
Even small clusters typically explore many disordered states before
they form locally crystalline structures.  The system needs to 
be very close to local equilibrium in order to ensure that the
ordered structures can be discriminated from the large number of
disordered states.  Thus, assembly is effective only when the
FDR is close to unity.  On the other hand, the directional interactions
in the model capsid system impose quite stringent constraints
on the local structure of the growing cluster.  This reduces the 
possibility for stable disordered clusters, 
and discriminating between ordered and disordered states is
easier.  Thus, the system still assembles effectively even when
deviations from local equilibrium are quite significant, and
assembly is still effective even when
deviations from FDT are quite large.

Taking account of these differences, we emphasize the main
feature of Fig.~\ref{fig:fdr}:
the FDR is large above the kinetic crossover, and small
below it.  We expect this behavior to be preserved 
as long three conditions are met.
Firstly, the observables used to construct the FDR
should couple to the processes
by which metastable disordered states are annealed into
ordered ones.  For example, if we had measured the FDR
in the capsid system using the capsomer
positions in place of their energies, then diffusive
processes would dominate both correlation and response
functions, and this response is not sensitive to the
extent to which the bonds in the system are locally
equilibrated.  

Secondly, we require that the assembled state of the system
minimizes the free energy both globally and locally.
Many biological systems are believed to have
funnel-shaped energy landscapes consistent with
this constraint \cite{funnel}.  The models 
presented in this article also have this property.  We believe
that satisfying this constraint contributes quite 
generally to good assembly, and it is therefore
practical to bear it in mind when designing self-assembling
systems.  Of course, systems that violate this
constraint do exist. 
For example, in three dimensions, minimization of the free
energy of small clusters of spherical 
particles lead to icosahedral structures \cite{icos},
while the crystalline phase has a close-packed structure.
It is therefore possible for these particles to assemble into 
icosahedral structures while always remaining locally equilibrated.
The FDR would be close to unity, but the system would  
never visit the `correctly assembled' close-packed
structure.  

The third condition that is required to ensure usefulness
of the parametric FDR plot concerns the time $t$ used
to construct it.
The behavior of Fig.~\ref{fig:fdr} depends weakly on the value of
$t$, but changing its order of magnitude will lead to different
behavior.  In particular, at very low temperatures
and for large $t$, the capsid system
shows an FDR close to unity.  This occurs because
the system is locally equilibrated over a particular set of disordered
states.  However, in this case,
the system would not be locally equilibrated while
the disordered clusters were forming, so that FDR
on that time scale would have been smaller than unity.  
In other words,
detection of the relevant deviations from local equilibrium requires
a measurement on the time scales during which those deviations occur.

These three conditions show that the application of the FDR 
to self-assembling systems requires some consideration
of the relevant observables and time scales.  However,
for the systems studied in this article, meeting these
conditions does not require careful tuning of model parameters,
but only the kind of qualitative analysis discussed in this 
section.  This represents evidence in favor of the applicability of 
these methods to other self-assembling systems.

\section{Outlook} 

The arguments of Section~\ref{sec:discussion} seem general,
and relatively independent of details of the system.
Further tests of the links between efficient
assembly, local equilibration, and FDRs would be valuable,
especially if FDRs could be measured experimentally.
In principle, FDRs can be obtained whenever conjugate 
correlation and response functions can be measured.  Measuring
fluctuations and responses of local quantities, such as the energy
of a single subunit, requires
a high degree of experimental control, but methods do exist
in some systems.  For example, Wang \emph{et al}~\cite{Wang06} recently
measured an FDR in a three dimensional glassy colloidal system.
The diffusive correlation function is conjugate to the response
of a single particle to a force in that case.  Applying similar
methods to ordering processes of spheres or discs
would be analogous to our studies of the sticky
disc system. 

Turning to biological systems,
it would be possible to measure the degree of kinetic frustration
in the folding of biomolecules, either computationally in more 
detailed capsid models, or in systems
such as the trpzip peptide~\cite{trpzip}, or experimentally,
in RNA folding, by a generalization of the experiment 
of~\cite{Liphardt_multiple}.  
In this latter case, the conjugate variables of force and displacement 
are already measurable, although obtaining good statistics
for the correlations and responses as a function of both $t$ and
$\tw$ might be challenging.  Results obtained in this way
would complement information about the non-equilibrium dynamics
obtained from analysis of the work distribution~\cite{Liphardt_jarz,
jarz_general}.
For example, the thermodynamic definitions of reversible and
irreversible work are linked to the idea that non-equilibrium
processes can occur with or without local equilibration.  
By characterizing the extent to which particular
degrees of freedom are locally equilibrated on particular
time scales, FDRs provide another link between these 
thermodynamic ideas and the statistical mechanics
of non-equilibrium trajectories. 

\begin{acknowledgments}
We thank Gavin Crooks, Ed Feng,
Juan Garrahan, Jan Liphardt, and Steve Whitelam 
for helpful discussions.  
RLJ was funded initially by NSF grant no. CHE-0543158 and later by the
Office of Naval Research Grant No. N00014-07-1-0689.  MFH was supported
initially by NIH Grant No. F32 GM073424-01, and later by the HHMI-NIBIB
Interfaces Initiative grant to Brandeis University.  DC was funded
initially by NSF grant no. CHE-0543158 and later by NSF grant no.
CHE-0626324.
\end{acknowledgments}

\begin{appendix}
\section{Time reversibility, and the FDR}

In this appendix, we briefly consider a general stochastic system 
evolving between times $t_\mathrm{i}$ and $t_\mathrm{f}$, and show
how deviations from the predictions of FDT come from 
trajectories (histories) which occur with probabilities that are 
different from those of their time reversed counterparts.

Consider a stochastic system evolving
between times $\ti$ and $\tf$.  The energy of 
a configuration $\C$ during this time period is given by
$E(\C)=E_0(\C) - hA(\C)$, where $h$ is a field, 
$A$ is an observable,
and $A(\C)$ its value in configuration $\C$.
The stochastic dynamics obey detailed balance with
respect to the Boltzmann distribution 
$p_\mathrm{eq}(\C)\propto e^{-\beta E(\C)}$.  
The response of observable $B$ to the field $h$ is
\begin{equation}
\chi(\tf,\ti) = \sum_{\C(t)} B(\Cf) 
\left.
\frac{\partial P[\C(t);h] }{\partial (\beta h)}\right|_{h=0}
\label{equ:resp_action}
\end{equation}
where the sum is over trajectories (histories) of the system, which
we indicate by the function $\C(t)$; the initial and final configurations
of the trajectory are $\Ci$ and $\Cf$ respectively; and
$P[\C(t);h]$ is the 
probability of a trajectory, which includes the probability
of its initial condition.

The property of detailed balance dictates that
\begin{eqnarray}
  \log \frac{P[\C(t);h]}{p_\mathrm{i}(\Ci)}  
- \log \frac{P[\Cbar(t);h]}{p_\mathrm{i}(\Cbari)}
& =& \beta h[ A(\Cf) - A(\Ci) ] \nonumber \\ & &
- \beta [ E(\Cf) - E(\Ci) ] \nonumber \\
\label{equ:det_bal}
\end{eqnarray}
where $p_\mathrm{i}(\Ci)$ is
the probability of the initial condition of the trajectory $\C(t)$,
and $\Cbar(t)$ is the time-reversed counterpart of 
$\C(t)$.  That is, $\Cbar(t) = 
\mathbb{T} \C(t_\mathrm{i}+t_\mathrm{f}-t)$, where the operator
$\mathbb{T}$ reverses all quantities that
are odd under time reversal, such as momenta.  To enforce
time-reversal symmetry of the equilibrium state, we assume that
the energy and its perturbation are time-reversal symmetric:
$E(\C)=E(\mathbb{T}\C)$ and $A(\C)=A(\mathbb{T}\C)$.  We 
also assume that $B(\C)=B(\mathbb{T}\C)$ for convenience,
although the same analysis can also be carried out without 
this assumption, leading to an analogous result.

Using $(\partial P[\C(t);h] /\partial h)
=P[\C(t);h] (\partial/\partial h) \log P[\C(t);h]$,
we substitute (\ref{equ:det_bal}) into (\ref{equ:resp_action}), and obtain
\begin{eqnarray}
\chi(\tf,\ti) &=& \langle B(\tf) [ A(\tf) - A(\ti) ] \rangle 
  + \nonumber \\ & &
 \sum_{\C(t)} B(\Cf) P[\C(t);0] 
\left. \frac{\partial}{\partial (\beta h)} \log P[\Cbar(t);h]  \right|_{h=0} 
\nonumber \\
\label{equ:fdr_chi1}
\end{eqnarray}
where we have used 
$\langle \cdot\rangle\equiv \sum_{\C(t)} P[\C(t)] (\cdot)$.
The fluctuation-dissipation theorem states that the first two terms
are equal at equilibrium, so we define $\Delta\chi(\tf,\ti) = 
\chi(\tf,\ti) - \langle B(\tf) [ A(\tf) - A(\ti) ] \rangle$ 
in order to measure deviations from FDT.

To obtain an informative 
expression for $\Delta\chi(\tf,\ti)$, we
use conservation of probability to write
\begin{equation}
\sum_{\C(t)} B(\Cbari) P[\Cbar(t);h] = \sum_{\C} B(\C) p_\mathrm{i}(\C)
\label{equ:tele}
\end{equation}
Thus, the derivative of the left
hand side of (\ref{equ:tele}) with respect to $h$ is zero. 
Noting that $B(\Cbari)=B(\Cf)$, we subtract this
derivative from the right hand side of (\ref{equ:fdr_chi1}), 
arriving at
\begin{eqnarray}
\Delta \chi(\tf,\ti) &=& 
 \sum_{\C(t)} B(\Cf) 
\left.\frac{\partial}{\partial (\beta h)} \log P[\Cbar(t);h] 
\right|_{h=0}
\times \nonumber \\ & & 
\qquad\qquad 
\{ P[\C(t);0] - P[\Cbar(t);0] \}
\label{equ:fdr_rev}
\end{eqnarray}
The purpose of (\ref{equ:fdr_rev}) is to show that if
all trajectories $\C(t)$ have the same probabilities as their
time-reversed counterparts $\Cbar(t)$, 
then the second term vanishes, and FDT applies.  This condition
holds exactly only at equilibrium, but if the dynamics
of the system are close to local equilibrium between times
$\ti$ and $\tf$, then the relative
weights of forward and reverse trajectories will be similar,
and deviations from FDT will be small.  

\end{appendix}

\end{document}